# Individual research performance: a proposal for comparing apples to oranges[1]

Giovanni Abramo

*National Research Council of Italy (IASI-CNR) and Laboratory for Studies of Research and Technology Transfer at University of Rome "Tor Vergata" – Italy*

ADDRESS: Dipartimento di Ingegneria dell'Impresa, Università degli Studi di Roma "Tor Vergata", Via del Politecnico 1, 00133 Roma – ITALY
tel. and fax +39 06 72597362, giovanni.abramo@uniroma2.it

Tindaro Cicero

*Laboratory for Studies of Research and Technology Transfer at University of Rome "Tor Vergata" – Italy*

ADDRESS: Dipartimento di Ingegneria dell'Impresa, Università degli Studi di Roma "Tor Vergata", Via del Politecnico 1, 00133 Roma – ITALY
tel. and fax +39 06 72595400, tindaro.cicero@uniroma2.it

Ciriaco Andrea D'Angelo

*Laboratory for Studies of Research and Technology Transfer at University of Rome "Tor Vergata" – Italy*

ADDRESS: Dipartimento di Ingegneria dell'Impresa, Università degli Studi di Roma "Tor Vergata", Via del Politecnico 1, 00133 Roma – ITALY
tel. and fax +39 06 72597362, dangelo@dii.uniroma2.it

---

# Individual research performance: a proposal for comparing apples to oranges

**Abstract**

The evaluation of performance at the individual level is of fundamental importance in informing management decisions. The literature provides various indicators and types of measures, however a problem that is still unresolved and little addressed is how to compare the performance of researchers working in different fields (apples to oranges). In this work we propose a solution, testing various scaling factors for the distributions of research productivity in 174 scientific fields. The analysis is based on the observation of scientific production by all Italian university researchers active in the hard sciences over the period 2004-2008, as indexed by the Web of Science. The most effective scaling factor is the average of the productivity distribution of researchers with productivity above zero.

# 1. Introduction

The evaluation of individual research performance is a fundamental tool for management, to inform decisions in areas such as faculty recruitment, career advancement, reward systems, grants awarding and projects funding. For the hard sciences, bibliometrics is particularly useful for large-scale measurement. Various performance indicators have been proposed, for use either singly or in combination. These range from indicators based on simple counts of publications over a time period, to those based on the scientist's overall number of citations (in some cases field-normalized), to the famous h-index (Hirsch, 2005) and its many variants, intended to synthesize both the quantity and quality of production. In their own work the authors have themselves developed and applied an indicator named "Fractional Scientific Strength", or FSS (Abramo and D'Angelo, 2011), which measures the field-normalized citations of a researcher's publications over the period of observation, while also accounting for fractional shares concerning joint-authored publications. While we cannot say that one particular indicator or combination is always the best, it is certainly possible to identify the one that is most appropriate given the objectives, conditions and context of the specific measurement task. The focus of this study is thus not to identify the best indicator for the measure of individual performance, rather to identify how to use them to compare researchers from different fields (apples to oranges). In the literature the authors have found no significant contributions on this topic. Scholars have mainly devoted efforts to the study of field-specific citation-density as a main topological property at the level of research groups (van Raan, 2008) or individuals (Costas et al., 2009). Because intensity of publication and citation behavior vary across research fields, responsible bibliometricians produce performance rankings by field (Sandström and Sandström, 2009; Zitt et al., 2005). Failure to do this can result in significant distortions in measurement and judgment (Abramo and D'Angelo, 2007). But as stated by Costas et al. (2009) "comparison of the statistical properties of the bibliometric performance indicator values of researchers and groups within the same population should be addressed in future work". In some way, this is a suggestion for the work that we are going to present.

In the Italian higher education system, which is the field of observation for our analysis, each faculty member is officially classified as belonging to a single field, known as a Scientific Disciplinary Sector (SDS). There are 370 SDSs grouped into 14 University Disciplinary Areas (UDAs). The problem we set is to compare the performance of scientists from different fields. One possibility is to do consider their percentile ranks of performance in their respective fields. Using this method we could compare the performance of a physicist to that of a mathematician, or within a single discipline that of a theoretical physicist to that of an experimental physicist. However the percentile rank presents certain limits (Thompson, 1993). Other than compressing the difference in performance between one position and the next, it is also sensitive to the size of the fields and the distribution of performance. For example, consider the case of a researcher from SDS A (with 10 researchers) who places third behind two researchers whose productivity is only a tiny bit better; his or her percentile rank will be 70. Then, consider another researcher from SDS B (with 100 researchers), who again places third with the same or an even greater gap: his or her percentile rank will be 97. The use of percentile ranks to compare the two researchers would certainly penalize the researcher from SDS A.

Thus in this work we wish to identify the most effective scaling factor for comparing research productivity among scientists from different fields. The utility of these findings is not limited to solving performance comparison between researchers in different fields, but extends most importantly to the problem of measuring and comparing research performance at the aggregate level of organizational units carrying out research in different fields or with unequal staffing for the respective fields, such as entire institutes, universities, departments, or even national education systems. In fact, for such unequal organizations, overall measures of performance can only be calculated by beginning from those of their individual researchers, appropriately standardized.

## 2. Methods

### 2.1 The productivity indicator FSS

Research activity is a production process in which the inputs consist of human, tangible (scientific instruments, materials, etc.) and intangible (accumulated knowledge, social networks, etc.) resources, and where outputs have a complex character of both tangible nature (publications, patents, conference presentations, databases, protocols, etc.) and intangible nature (tacit knowledge, consulting activity, etc.). The new-knowledge production function therefore has a multi-input and multi-output character. The principal efficiency indicator of any production system is labor productivity. To calculate it we need to adopt several simplifications and assumptions. The first simplification is to consider the publications indexed in the Web of Sciences (WoS) as a proxy of total output. The second assumption is that the resources available to all researchers within the same field of observation are the same. The third assumption is that the hours devoted to research are more or less the same for all professors.

A very gross way to calculate the average yearly labor research productivity is to simply measure the fractional count of publications per researcher in the period of observation and divide it by the full time equivalent of labor in the period. A more sophisticated way to calculate productivity recognizes the fact that publications, embedding the new knowledge produced, have different values. Their value depends on their impact on scientific advancements. As proxy of impact, bibliometricians adopt the number of citations for the researchers' publications, in spite of the potential distortions inherent in this indicator (Glänzel, 2008). However, comparing researchers' performance by field is not enough to avoid distortions in rankings. In fact citation behavior also varies across fields, and it is not unlikely that researchers belonging to a particular scientific field may further publish outside their field. For this reason we standardize the citations for each publication with respect to the average for the distribution of citations for all the Italian cited publications of the same year and the same subject category (Abramo, Cicero, & D'Angelo, 2012).

Our best proxy of the average yearly productivity at the individual level is the indicator named FSS:

$$FSS = \frac{1}{t} \cdot \sum_{i=1}^{N} \frac{c_i}{\bar{c}_i} * f_i$$

Where:

t = number of years of work of the researcher in the period of observation (2004-2008);
N = number of publications of the researcher in the period of observation;
$c_i$ = citations received by publication *i* (accumulated at June 30, 2009*)*;
$\bar{c}_i$ = average of the distribution of citations received for all Italian cited publications of the same year and subject category of publication *i*;
$f_i$ = fractional contribution of the researcher to publication *i*.

In fields where the author list is given in alphabetical order, the fractional contribution for each author equals the inverse of the total number. In other cases, particularly in the life sciences, widespread practice is to indicate the various contributions to the published research by the positioning of the names in the byline. For the life sciences then, when the number of co-authors is above two, different weights are given to each co-author according to their position in the list and the character of the co-authorship (intra-mural or extra-mural). If first and last authors belong to the same university, 40% of citations are attributed to each of them; the remaining 20% are divided among all other authors. If the first two and last two authors belong to different universities, 30% of citations are attributed to first and last authors; 15% of citations are attributed to second and last author but one; the remaining 10% are divided among all others[2].

In the Italian higher education system all professors of the same academic rank receive the same stipend, with variation by seniority. Since information on individuals is subject to privacy, for the present work it seems sufficient to approximate a professor's salary as being the average for his or her academic rank. Thus to standardize FSS for the salaries and take into account the varying unit costs of labor, we calculate the relative coefficients as seen in Table 1 and then normalize the average yearly productivity of each professor to a stipend coefficient, which represents the ratio of the professor's stipend to that of assistant professors:

$$FSS^* = \frac{FSS}{Stipend\ coefficient}$$

*Table 1: Yearly average stipend of 2004-2008 professors per academic rank (source DALIA - https://dalia.cineca.it/php4/inizio_access_cnvsu.php)*

| Academic rank | Yearly average stipend (euro) | Stipend coefficient |
|---|---|---|
| Full professor (confirmed)* | 124,939 | 2,783 |
| Full professor (probationary) | 94,442 | 2,103 |
| Associate professor (confirmed) | 90,622 | 2,018 |
| Associate professor (probationary) | 68,469 | 1,525 |
| Assistant professor (confirmed) | 68,844 | 1,533 |
| Assistant professor (probationary) | 44,899 | 1 |
| Research assistant (obsolete rank) | 81,721 | 1,820 |

---

[2] The weighting values were assigned following advice from senior Italian professors in the life sciences. The values could be changed to suit different practices in other national contexts.

Should information on salary be unavailable, the second best proxy is to calculate productivity rankings per academic rank.

**2.2 Potential scaling factors for standardizing individual productivity**

In our analysis we consider four potential different scaling factors for the purposes of comparing researchers working in different scientific fields:

1. $FSS_a^* = \frac{FSS^*}{average}$: ratio of FSS[*] to the average value of the distribution of FSS[*] for the same SDS;
2. $FSS_{a0}^* = \frac{FSS^*}{average\ (no\ zero\ FSS^*)}$: variant of the preceding indicator, with the difference that scientists with nil values of FSS[*] are not included in calculation of the average;
3. $FSS_m^* = \frac{FSS^*}{median}$: ratio between the value of FSS[*] for a scientist and the median value of the distribution of FSS[*] for the same SDS and period;
4. $FSS_{m0}^* = \frac{FSS^*}{median\ (no\ zero\ FSS^*)}$: variant of the preceding indicator, with the difference that scientists with nil values of FSS[*] are not included in calculation of the median.

**2.3 Dataset**

The dataset of Italian professors active in the hard sciences over the period 2004-2008 is extracted from an official database maintained by the Ministry of Universities and Research (http://cercauniversita.cineca.it/php5/docenti/cerca.php). The ministry database provides the information on every professor's home university, SDS and academic rank. The information on the professors' scientific publications is extracted from the Italian Observatory of Public Research (ORP - www.orp.researchvalue.it), a database developed by the authors and derived under license from the raw data of the Thomson-Reuters Italian National Citation Report, an extract of the WoS. The ORP employs a complex algorithm to resolve ambiguity in author names and home institutions, permitting a reduction to 4% error (harmonic average of precision and recall) in identifying each professor with their proper WoS publications (D'Angelo, Giuffrida, & Abramo, 2011). To render the bibliometric analysis still more robust, the field of observation is limited to those SDSs in the hard sciences (174 out of total 205) where at least 50% of the member academics produce at least one publication with a citation[3] in the 2004-2008 period. This is a total of 38,700 scientists.

In the subsequent sections of this study we adopt an analytical method involving graphic presentations of the distributions of productivity under the different scaling factors. For a clearer presentation we provide only the analyses for the subset of SDSs including the largest and the smallest (with at least 40 professors) SDSs in each UDA, namely:

---

[3] In order to avoid SDSs with a median productivity equal to nil.

- Mathematical Analysis (MAT/05) and Mathematical logic (MAT/01);
- Experimental Physics (FIS/01) and Physics for Earth and Atmospheric Sciences (FIS/06);
- Organic chemistry (CHIM/06) and Environmental Chemistry and Chemistry for Cultural Heritage (CHIM/12);
- Stratigraphic and Sedimentological Geology (GEO/02) and Applied Geophysics (GEO/11);
- Biochemistry (BIO/10) and Anthopology (BIO/08);
- Internal Medicine (MED/09) and Neuroradiology (MED/37);
- Agronomy and Herbaceous Cultivation (AGR/02) and Animal Husbandry (AGR/20);
- Construction Science (ICAR/08) and Environmental and Health Engineering (ICAR/03);
- Data Processing Systems (ING-INF/05) and Theory of Development for Chemical Processes (ING-IND/26).

Table 2 shows the sizes of the overall UDAs and of the SDSs chosen for the analysis.

*Table 2: Dimensions and selection of SDSs for the analysis*

| UDA | n. of SDSs | Research staff | SDS for analysis | Share of the UDA research staff |
|---|---|---|---|---|
| Mathematics and Computer Sciences | 9 | 3,514 | MAT/05 | 0.268 |
|  |  |  | MAT/01 | 0.012 |
| Physics | 7 | 2,822 | FIS/01 | 0.387 |
|  |  |  | FIS/06 | 0.024 |
| Chemistry | 12 | 3,605 | CHIM/06 | 0.206 |
|  |  |  | CHIM/12 | 0.020 |
| Earth Sciences | 12 | 1,438 | GEO/02 | 0.150 |
|  |  |  | GEO/11 | 0.042 |
| Biology | 19 | 5,777 | BIO/10 | 0.172 |
|  |  |  | BIO/08 | 0.016 |
| Medicine | 47 | 12,199 | MED/09 | 0.100 |
|  |  |  | MED/37 | 0.004 |
| Agricultural and veterinary sciences | 25 | 2,894 | AGR/02 | 0.071 |
|  |  |  | AGR/20 | 0.019 |
| Civil engineering | 6 | 1,341 | ICAR/08 | 0.280 |
|  |  |  | ICAR/03 | 0.071 |
| Industrial and information engineering | 37 | 5,110 | ING-INF/05 | 0.132 |
|  |  |  | ING-IND/26 | 0.008 |
| Total | 174 | 38,700 |  | 0.182 |

## 3. Results and analysis

### 3.1 Distribution of individual productivity

As a first step, we calculate descriptive statistics for the individual productivity in each SDS (Table 3). In keeping with the observations in the literature, there are evident differences among the SDSs. The SDS with the highest average productivity (0.313) is CHIM/06, and the one with the lowest average (0.053) is BIO/08, respectively from the Chemistry and Biology UDAs. These particular SDSs also present remarkably different asymmetry in their productivity distributions, with CHIM/06 being much more accentuated (skewness 12.076), indicating that here there are more professors with productivity far above the average than occur in BIO/08.

The other 16 SDSs show average values of productivity varying between 0.080 in MED/37 and 0.232 in ING-IND/26. The SDS with the most uniform distribution around its average is MED/37 (coefficient of variation 127.6). The differences among the SDSs are also clearly visible in examining the incidence of scientists with nil FSS*. The CHIM/06 and BIO/10 SDSs have quite low percentages of unproductive professors, respectively at 4.2% and 7.9%. On the opposite front, the SDS with the highest incidence of unproductive professors is MAT/01 (39.0%), while there are also very high percentages in AGR/02 (37.9%), BIO/08 (33.3%) and MAT/05 (30.7%).

Besides characterizing the differences through descriptive statistics, we also provide a graphic representation of the distributions of productivity in the 18 chosen SDSs. To do this we adopt a method of representing the distribution that permits the reader to observe if the direct comparison between researchers in different SDSs is plausible. On a log scale, we diagram the probability (Y axis) of observing a value of productivity greater than or equal to a given value (in the X axis). In Figure 1 we observe that the probability functions of the 18 SDSs are not directly comparable, in that the probability curves cannot be superimposed.

*Table 3: Descriptive statistics of FSS\* per SDS*

| SDS | Professors | With nil FSS* (%) | Average | Variat. coeff. | Median | IQR† | Skewness |
|---|---|---|---|---|---|---|---|
| AGR/20 | 55 | 12.7% | 0.088 | 122.4 | 0.046 | 0.101 | 1.897 |
| AGR/02 | 206 | 37.9% | 0.081 | 199.8 | 0.012 | 0.114 | 4.191 |
| BIO/08 | 90 | 33.3% | 0.053 | 202.9 | 0.008 | 0.056 | 3.445 |
| BIO/10 | 991 | 7.9% | 0.155 | 179.6 | 0.071 | 0.150 | 6.691 |
| CHIM/12 | 73 | 13.7% | 0.177 | 169.5 | 0.082 | 0.209 | 4.436 |
| CHIM/06 | 742 | 4.2% | 0.313 | 205.9 | 0.169 | 0.299 | 12.076 |
| FIS/06 | 68 | 11.8% | 0.155 | 243.7 | 0.055 | 0.110 | 6.134 |
| FIS/01 | 1114 | 11.1% | 0.170 | 147.7 | 0.080 | 0.192 | 3.671 |
| GEO/11 | 61 | 26.2% | 0.086 | 178.9 | 0.028 | 0.094 | 2.819 |
| GEO/02 | 216 | 21.3% | 0.082 | 151.1 | 0.040 | 0.097 | 2.935 |
| ICAR/03 | 95 | 27.4% | 0.103 | 141.7 | 0.045 | 0.131 | 1.928 |
| ICAR/08 | 375 | 29.6% | 0.161 | 210.5 | 0.046 | 0.183 | 6.060 |
| ING-IND/26 | 43 | 16.3% | 0.232 | 160.8 | 0.061 | 0.268 | 2.500 |
| ING-INF/05 | 673 | 18.9% | 0.180 | 176.9 | 0.070 | 0.212 | 5.330 |
| MAT/01 | 41 | 39.0% | 0.119 | 192.5 | 0.027 | 0.103 | 2.619 |
| MAT/05 | 942 | 30.7% | 0.177 | 185.1 | 0.052 | 0.212 | 4.050 |
| MED/37 | 46 | 13.0% | 0.080 | 127.6 | 0.040 | 0.107 | 1.687 |
| MED/09 | 1224 | 17.6% | 0.208 | 182.9 | 0.061 | 0.245 | 4.409 |

† *Interquartile range*

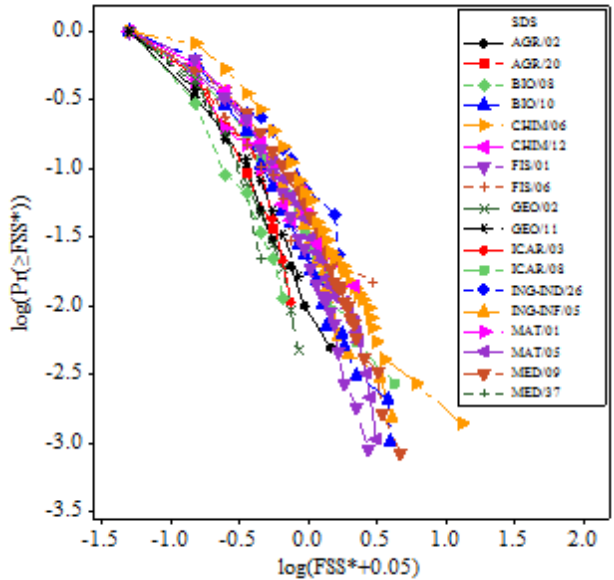

*Figure 1: Graphic representation of FSS\* for the analyzed SDSs (the +0.05 on the X-axis allows inclusion of non-productive scientists - i.e. with FSS = 0 - in the distribution plots).*

Taking these distributions, we now attempt to reduce them to a single function that could provide a general theoretical model to explain the trends for the distributions of all the fields analyzed. Given the productivity measure being based on positive asymmetric distributions of citations, we try to fit it with a generalized Pareto distribution. For significance reasons, we run the test for the largest SDS only, in each of the nine UDAs: we estimate the three Pareto function parameters and measure the goodness of fit by the Kolmogorov-Smirnov test. Table 4 shows the results obtained. In none of the 9 SDSs analyzed, the generalized Pareto distribution provides an optimal fit[4]: all the values of the Kolmogorov-Smirnov test are above the 5% critical value, consequently we have to reject the hypothesis of goodness of fit of a Pareto distribution. In light of these results indicating that we cannot use a single theoretical model to compare the distributions for different UDAs, we now depict the empirical frequency distributions and test the effectiveness of the four methods of standardization that we initially suggested.

---

[4] In this case the null hypothesis corresponds to a good fit of data, so in rejecting it (chi-square values stars) we indicate the failure to obtain good fit.

*Table 4: Statistics of fit to a generalized Pareto distribution for the largest SDSs in each UDA*

| SDS | UDA | Generalized Pareto parameters | | | | Goodness of fit | |
|---|---|---|---|---|---|---|---|
| | | Obs. | k | Sigma | μ | K-S test* | Critical value (5%) |
| MAT/05 | Mathematics and Computer Sciences | 942 | 0.436 | 0.113 | -0.024 | 0.180 | 0.004 |
| FIS/01 | Physics | 1114 | 0.290 | 0.129 | -0.012 | 0.089 | 0.041 |
| CHIM/06 | Chemistry | 742 | 0.360 | 0.199 | 0.003 | 0.051 | 0.050 |
| GEO/02 | Earth Sciences | 216 | 0.308 | 0.064 | -0.010 | 0.140 | 0.092 |
| BIO/10 | Biology | 991 | 0.368 | 0.103 | -0.008 | 0.072 | 0.043 |
| MED/09 | Medicine | 1224 | 0.427 | 0.135 | 0.027 | 0.173 | 0.039 |
| AGR/02 | Agricultural and veterinary sciences | 206 | 0.501 | 0.047 | -0.012 | 0.213 | 0.095 |
| ICAR/08 | Civil engineering | 375 | 0.490 | 0.092 | -0.019 | 0.181 | 0.070 |
| ING-INF/05 | Industrial and information engineering | 673 | 0.371 | 0.126 | -0.020 | 0.142 | 0.052 |

*\* Kolmogorov-Smirnov test*

### 3.2 Analysis of standardized distributions

As a first case, we analyze the effects of standardization by the average and median calculated over the entire distribution of productivity (including nil values) for each SDS.

Figure 2 presents the standardization to average value ($FSS_a^*$) in a log-scale, for a better visualization[5]. What we observe is a superimposition of the probability curves in the central part of the distributions. There is less comparability in the extreme values, particularly for the best scientists, identifiable in the graph as those with a log of $FSS_a^*$ greater than 2.

Standardization to the median (Figure 3) amplifies the differences among the probability curves. We observe that particularly for the SDSs with strongly different concentrations of professors with nil productivity, such as AGR/02 (37.9%) and CHIM/06 (4.2%), the probability curves are remarkably divergent. The median is clearly very sensitive to the massive presence of nil values seen in AGR/02.

Moving on to the standardizations that exclude the nil values from the productivity distributions (Figure 4), we observe that again in this case the average is an effective scaling factor for the central part of the distribution and slighly less so for the tails.

The graph shows a slightly greater superimposition of the curves compared to the previous standardization to the average, where we also included the researchers with nil productivity. The best possibility for comparison among fields is in the interquartile range, for values of productivity in the second and third quartiles. We observe also that the larger the size of the SDSs, the more effective the scaling by the average excluding nil values (Figure 5).

The results are less encouraging considering standardization to the median, again calculated with exclusion of nil productivity values (Figure 6).

---

[5] Because probability values (Y axis) are all below 1, they are negative in the log-scale. The same occurs for FSS values below 1 (X axis).

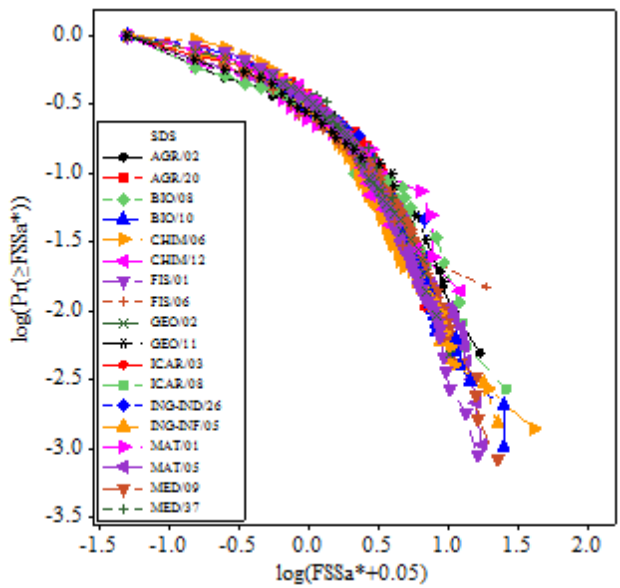

*Figure 2: Graphic representation of productivity standardized to average for the SDSs examined, including nil values (unproductives researchers)*

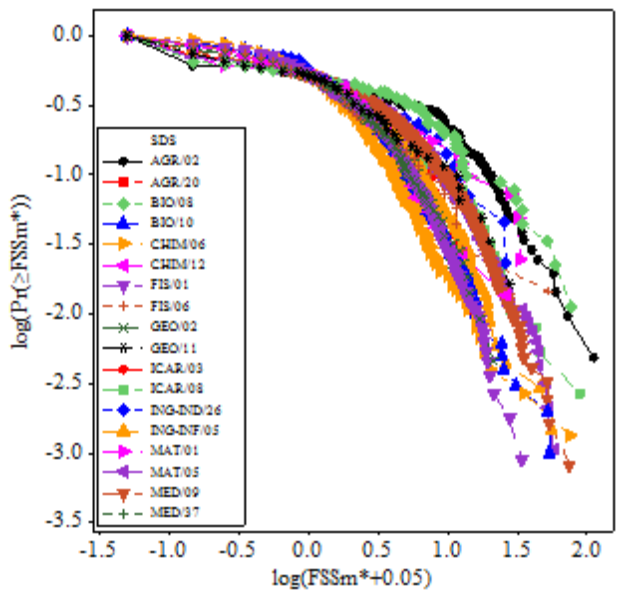

*Figure 3: Graphic representation of productivity standardized to the median for the SDSs examined, including nil values*

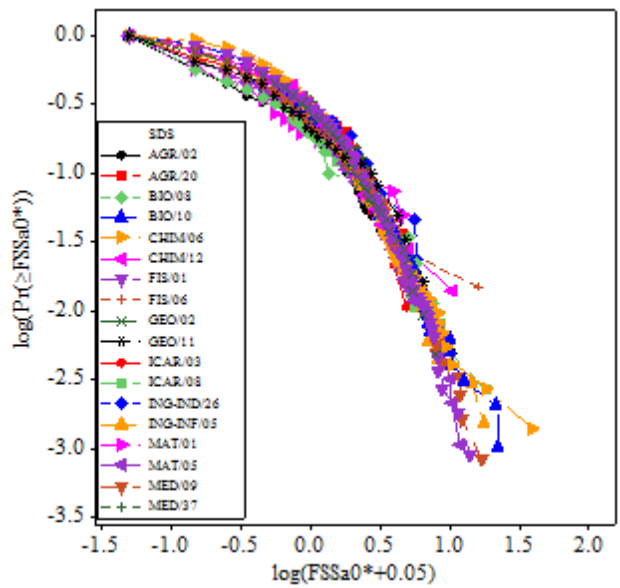

*Figure 4: Graphic representation of productivity standardized to the average for the SDSs considered, excluding nil values*

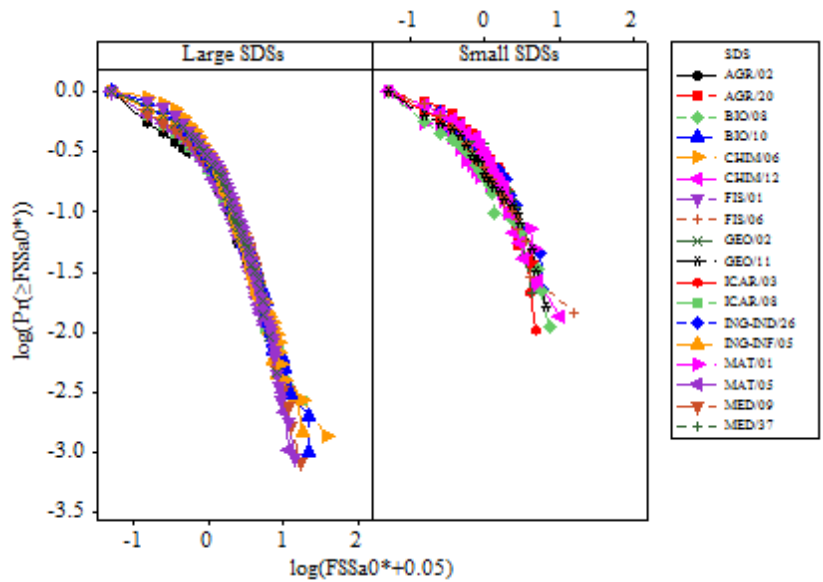

*Figure 5: Comparison between large and small SSDs, standardizing to the average of FSS\*, excluding nil values.*

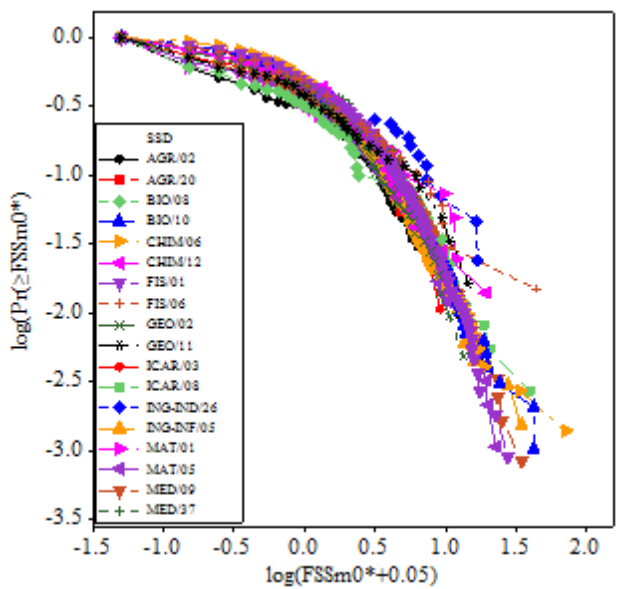

*Figure 6: Graphic representation of productivity standardized to the median for the SDSs considered, excluding nil values*

Overall, the average excluding nil values is the best scaling factor for standardization of individual values of productivity based on citation counts. Still, in comparing the probability curves for all four scaling factors tested, there is a divergence for the extreme values of the distributions. For this reason we now turn our focus specifically to these extreme values with two in-depth analyses: one concerning the top scientists and one concerning the outliers.

**3.3 Analysis of extreme values**

*3.3.1 Analysis of top scientists*

To analyze the behavior of the scaling factors for extreme values, our first step is to construct a global researcher productivity ranking for the 174 SDSs. We then extract the researchers that belong to the top 5%, 10% and 20% of these global rankings and carry out various simulations. If the standardization is effective and thus the distributions of the various SDSs are comparable to each other, then we expect that the percentages of top scientists within each SDS will be the same. Admissible variation of this percentage is measured by standard deviation, in keeping with (Radicchi and Castellano, 2011), as shown in Table 5.

The first analysis we present is for the extraction of the global top 5%, as seen in Figure 7. The figure shows the percentage of scientists belonging to the top 5% of the global ranking by SDS for each type of standardization. Each bar represents one of the 18 SDSs analyzed and the dashed lines represent admissible variations for 1 standard deviation.

*Table 5: Admissible variation in percentage for classes of top scientists*

| | | average±1std dev. | |
|---|---|---|---|
| Expected value | Std deviation | Lower bound | Upper bound |
| 5% | 2.32% | 2.7% | 7.3% |
| 10% | 3.20% | 6.8% | 13.2% |
| 20% | 4.26% | 15.7% | 24.3% |

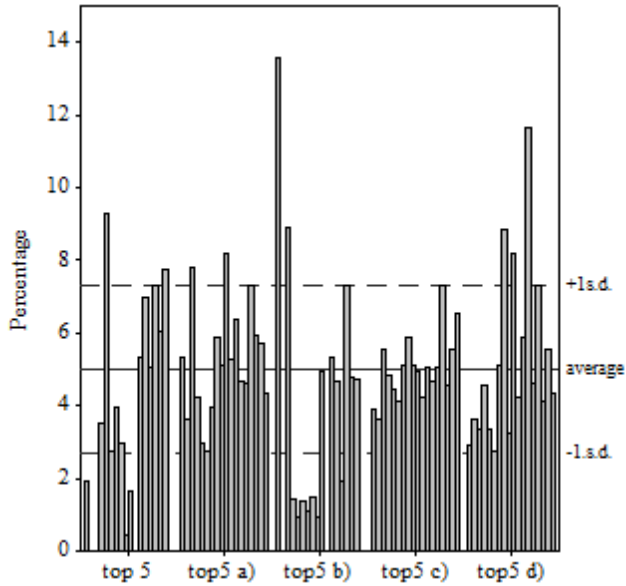

*Figure 7: Percentage of scientists belonging to the top 5% of the global ranking by SDSs and type of standardization – (a) Standardization to average with nil values; (b) Standardization to median with nil values; (c) Standardization to average without nil values; (d) Standardization to median without nil values.*

We observe that in the absence of standardization (the first group of bars at left) the scientists belonging to SDS CHIM/06 (third from left) would gain notable advantage, arriving at a 9.3% share of globally-ranked top scientists, a value that exceeds the upper bound of average ± 1 std dev. (7.30%).

Moving to the right in Figure 7, the sequence of four blocks of bars represents the four methods of standardization proposed: a) standardization to the average of the entire distribution; b) to the median of the entire distribution; c) to the average net of nil values; d) to the median net of nil values.

We observe that in the case of standardizing to the average net of nil productivity values, all SDSs show a percentage of top 5% that falls in the average± 1 std dev. interval, indicating that this is the most reliable scaling factor. The standardization methodology that gives rise to the greatest oscillation is the one for median calculated over the entire distribution, significantly distorting the percentage of top 5%. In AGR/02 – the first bar on the left of this group – the median of productivity is quite low and the percentage of top scientists arrives at 13.6% of total. A separate analysis for the nine largest and nine smallest SDSs shows that the scaling is more effective for the largest ones (Figure 8).

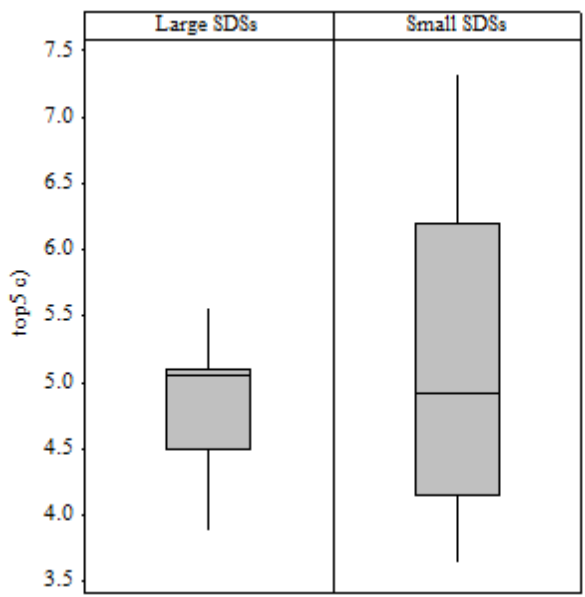

*Figure 8: Box plots of distribution of percentage of top 5% by SSDs size.*

We repeat the analysis extracting the top 10% scientists (Figure 9). The results reveal that scaling by the average obtained excluding nil values is not always better than the overall average. Finally, if we consider the distribution of top 20% scientists, it is evident that in all SDSs the average including nil values is the best scaling factor (Figure 10).

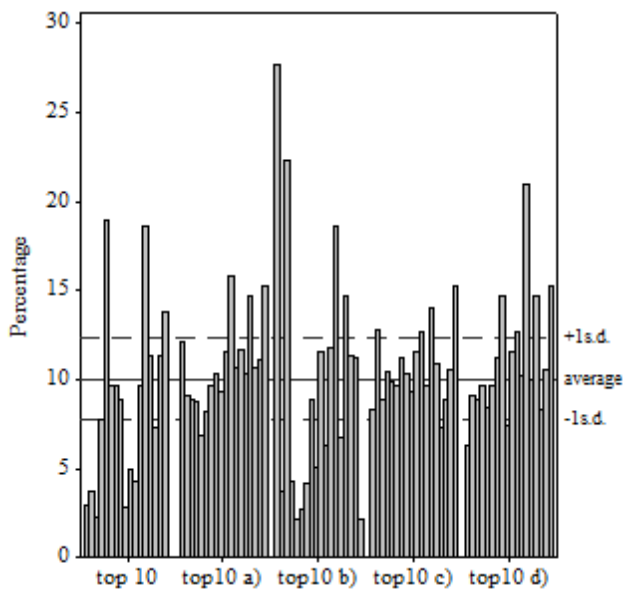

*Figure 9: Percentage of scientists belonging to the top 10% of the global ranking by SDSs and type of standardization – (a) Standardization to average with nil values; (b) Standardization to median with nil values; (c) Standardization to average without nil values; (d) Standardization to median without nil values.*

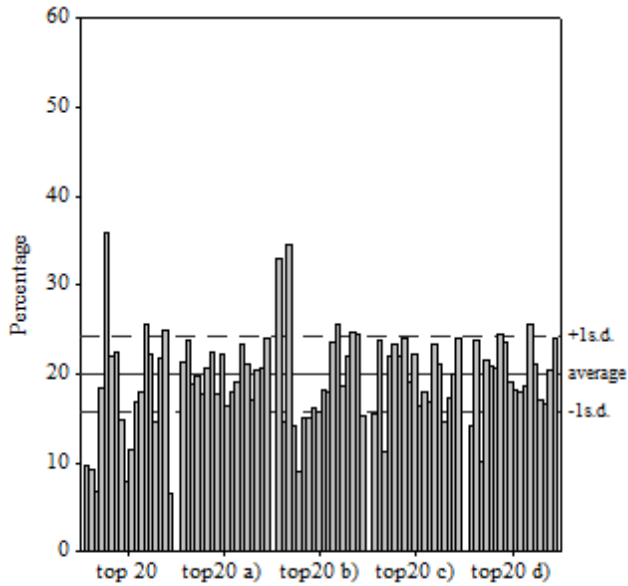

*Figure 10: Percentage of scientists belonging to the top 20% of the global ranking by SDSs and type of standardization – (a) Standardization to average with nil values; (b) Standardization to median with nil values; (c) Standardization to average without nil values; (d) Standardization to median without nil values.*

*3.3.2 Analysis of outliers*

The above analysis on extreme values for top scientists does not provide definitive answers on the most effective scaling factor. Therefore, in this last section we further attempt to evaluate the effects of the various standardizations on the "outlier" values of performance. These values are responsible for considerable distortions in aggregate productivity measures. We begin by constructing a single probability distribution without distinction by SDS, eliminating the nil productivity values so as to avoid a distribution centered so low that it would create false positives (values that are not outliers but appear to be so).

To identify the outliers we then use a non-parametric "outlier detection algorithm" known as MAD (median of the absolute deviation). The technique requires identifying the median value of the distribution and then calculating the absolute difference (D) between each productivity value and the median. Taking the distribution for D we again calculate the median, which now takes the name MAD. At this point a performance value is considered an outlier if $\frac{|D|}{MAD} > 5$ (Sprent and Smeeton, 2001). Out of the non-standardized values of productivity (a total of 5,830 observations for the professors with "non-nil" productivity in the 18 SDSs considered), there are 633 outliers (10.86%). The incidence of these outlying performers in the total research staff of the SDSs is shown in Table 6, Column 2. The percentages vary from 0 in AGR/20 and MED/37, to 22.2% for ING-IND/26, for a total range of 22.2 points. Standardization to the average without nil values generates a reduction in outliers, down to 566 from 633, but more importantly reduces the unevenness in incidence of outliers among the SDSs: from the previous 22.2 points, the range of variation drops to 6.8, from a minimum of 7.7 in CHIM/06 to the maximum in ICAR/03 at 14.5.

*Table 6: Percentage of outliers and their distribution per SDS under different scaling factors*

| SDSs | $FSS^*$ | $FSS^*_a$ | $FSS^*_m$ | $FSS^*_{a0}$ | $FSS^*_{m0}$ |
|---|---|---|---|---|---|
| AGR/02 | 3.9 | 15.6 | 45.3 | 9.4 | 7.8 |
| AGR/20 | 0.0 | 10.4 | 4.2 | 10.4 | 6.3 |
| BIO/08 | 1.7 | 13.3 | 33.3 | 11.7 | 11.7 |
| BIO/10 | 6.4 | 7.8 | 5.1 | 9.1 | 8.9 |
| CHIM/06 | 16.6 | 6.5 | 2.5 | 7.7 | 6.9 |
| CHIM/12 | 6.3 | 6.3 | 4.8 | 7.9 | 6.3 |
| FIS/01 | 8.7 | 8.6 | 5.5 | 9.6 | 9.8 |
| FIS/06 | 10.0 | 10.0 | 10.0 | 10.0 | 15.0 |
| GEO/02 | 3.5 | 10.0 | 7.1 | 10.0 | 7.1 |
| GEO/11 | 6.7 | 15.6 | 15.6 | 13.3 | 15.6 |
| ICAR/03 | 5.8 | 15.9 | 10.1 | 14.5 | 13.0 |
| ICAR/08 | 12.5 | 13.6 | 16.7 | 11.0 | 12.5 |
| ING-IND/26 | 22.2 | 11.1 | 22.2 | 13.9 | 22.2 |
| ING-INF/05 | 10.4 | 9.5 | 9.2 | 9.5 | 9.5 |
| MAT/01 | 12.0 | 16.0 | 24.0 | 12.0 | 16.0 |
| MAT/05 | 14.2 | 12.9 | 18.7 | 10.0 | 9.3 |
| MED/09 | 14.7 | 11.5 | 14.7 | 10.5 | 10.7 |
| MED/37 | 0.0 | 12.5 | 5.0 | 12.5 | 12.5 |
| Number of outliers | 633 | 581 | 614 | 566 | 559 |

## 4. Conclusions

Management of research requires robust and reliable quantitative information to evaluate individual researchers and support decisions in areas such as faculty recruitment, career advancement, reward programs, grant decisions and project funding. The literature provides a variety of performance indicators for the purposes, however little attention has been given to exploring their applicability for comparing researchers working in different fields. This study has attempted to provide a contribution.

Based on the scientific production by Italian academics as indexed in the Web of Science for the period 2004-2008, we analyzed and compared the distributions of productivity in the largest scientific disciplinary sectors of each of the nine hard science disciplines. From the analysis, it emerges that individual productivity in research tends to be distributed in a manner that is completely different from one SDS to the next. There are only three out of nine SDSs where the productivity distribution fits a generalized Pareto distribution (a distribution skewed to the right), and for the remaining two thirds of the SDSs it was not possible to find a distribution function that could represent them and so permit ready comparison. Thus, to arrive at a robust comparison of empirical productivity values for researchers belonging to different SDSs, we tested four scaling factors, finding that the most effective one is the average of the distribution of researchers with productivity above zero. This result was confirmed by two analyses of the right tails of the distributions, which represent the critical cases of top scientists and outliers. The average excluding nil values succeeds in appropriately rescaling the productivity values for very top scientists in general and outliers in particular, making it possible to obtain an objective comparison of their performance, independent of the SDSs to which they belong.

The authors are convinced that the results can provide concrete assistance to evaluators, assisting them to avoid errors in comparing the research performance of "apples to oranges". Apart from individual researchers, the result is also of fundamental importance for measuring and comparing performance at the aggregate level of organizational units constructed of unlike fields and with unequal staffing.